# Stabilization Mechanism of ZnO nanoparticles by Fe doping

Jianping Xiao[1,2], Agnieszka Kuc[1], Thomas Frauenheim[2], Thomas Heine[1,*]
*1 School of Engineering and Science, Jacobs University Bremen, Campus Ring 1, 28759 Bremen, Germany,
2 Bremen Center for Computational Materials Science, Universität Bremen, Am Fallturm 1, 28359 Bremen, Germany*

Surprisingly low solubility and toxicity of Fe-doped ZnO nanoparticles is elucidated on the basis of first-principles calculations. Various ZnO surfaces that could be present in nanoparticles are subject to substitutional Fe doping. We show that Fe stabilizes polar instable surfaces, while non-polar surfaces, namely (10$\underline{1}$0) and (11$\underline{2}$0), remain intact. Polar surfaces can be stabilized indirectly through $Fe^{2+}$-$Fe^{3+}$ pair assisted charge transfer, what reduces surface polarity and therefore, the solubility in polar solvents.

Due to its promising electronic and optoelectronic properties, zinc oxide (ZnO) has been increasingly investigated and applied in many fields, such as photocatalysis [1], optical devices [2], and cosmetic products, such as sunscreens [3]. However, ZnO nanoparticles (NPs) significantly differ in chemical and physical properties from the macrosized bulk material with identical chemical composition. ZnO nanoparticles have been reported to be toxic [4], and several research groups have demonstrated its negative effects on living organisms and tissues under cellular environment [5]. Though similar in size, ZnO NPs are significantly more toxic than $TiO_2$ and $CeO_2$ NPs due to solvated $Zn^{2+}$ ions [6]. The solubility, release of toxic $Zn^{2+}$ ions, and thus, the toxicity of ZnO NPs can be, however, strongly suppressed by substitutional doping, *e.g.* with iron [7].

In our previous work, we have investigated the influence of Fe doping of ZnO NPs (Fe-ZnO NPs) on the structure, energetic stability, and also the electronic properties. We showed that substitutional $Fe^{2+}$ doping of up to 12.5 at% leaves the ZnO wurtzite lattice intact [8]. We have also supported the experimental findings that Fe dopants stabilize ZnO NPs and distribute homogeneously inside the ZnO lattice. Moreover, based on the experimental and theoretical inner-shell and Mößbauer spectroscopy, we showed that Fe dopants occur in ZnO NPs predominantly with oxidation state $Fe^{2+}$ [8,9].

The reduction of ZnO NP toxicity has been recognized to be related to the suppression of $Zn^{2+}$ ion release to the solution, and the solvation process most probably will take place at the surface of the ZnO NPs [10]. The Fe doped ZnO surfaces and the stabilization processes of such mixed materials are highly unexplored and therefore, we study here the surface stabilization mechanism in these Fe-ZnO NPs. Our results show that ZnO surfaces are stabilized by Fe substitutional doping and this process takes place through $Fe^{2+}$-$Fe^{3+}$ pair assisted charge transfer, leading to the decrease in the polarity of polar (0001) surfaces.

ZnO possesses two types of dominant surfaces: non-polar (10$\underline{1}$0) and (11$\underline{2}$0) surfaces, and polar (0001) surfaces, terminated either by Zn, (0001)-Zn, or O, (000$\underline{1}$)-O [11,12]. Contrary to the non-polar surfaces, the polar surfaces exhibit a significant dipole moment perpendicular to the surface. The total dipole moment of ZnO bulk is accumulated along the crystallographic [0001] direction. Meanwhile, the presence of a finite dipole moment per surface area for polar surfaces gives rise to a great macroscopic electrostatic field, which scales with the thickness of the ZnO crystals. Hence, the polar surfaces are always energetically the least stable ones, what is known as the "polar instability problem" [13]. Therefore, in order to reduce solubility of ZnO, one needs to reduce the polarity of its surfaces.

There are several scenarios to suppress the dipole moment and thus to stabilize polar surfaces. One of them follows a partial charge transfer from the oxygen-terminated surface to the zinc-terminated one without significant surface reconstruction. Another way is correlated with a reconstruction of polar surfaces, accompanied by formation of zinc and oxygen vacancies. Adsorption of charged adparticles is yet another stabilizing factor [14,15]. Certainly, any combination of the above-mentioned scenarios is also possible. First principles calculations have already confirmed that adatoms and surface reconstruction have a stabilizing effect on the polar surfaces [16,17]. However, there are several experimental works reporting that polar ZnO surfaces, in fact, do not attract adparticles [18,19] and they undergo a variety of surface reconstructions [20,21]. Therefore, the ZnO surface stabilization due to Fe dopants is still not understood.

We have studied the behaviour of Fe dopants on the dominant and clean non-polar and polar ZnO surfaces. In addition, defective polar surfaces with zinc and oxygen vacancies were considered. Furthermore, we have studied the oxidation states of Fe placed at different positions of polar surfaces and the charge transfer in-between two polar surfaces in order to understand the relevant stabilization mechanism of Fe-ZnO surfaces. Finally, we have simulated the X-ray Absorption Near Edge Structure (XANES) to provide reference data for a surface-sensitive characterization method.

All studied ZnO surfaces have been simulated using two-dimensional periodic boundary conditions. In order to reflect low Fe concentrations we have employed the supercell approach, using (2×1) supercell for the non-polar (10$\underline{1}$0) surfaces (containing 64 atoms) and (2×2) supercells for the polar (0001)-O and (0001)-Zn surfaces (containing 72 atoms) as shown in Fig. 1. Hereafter, Fe@$N$ refers to the Fe dopants placed at the substitutional Zn site in the atomic layer $N = a, b, c$.

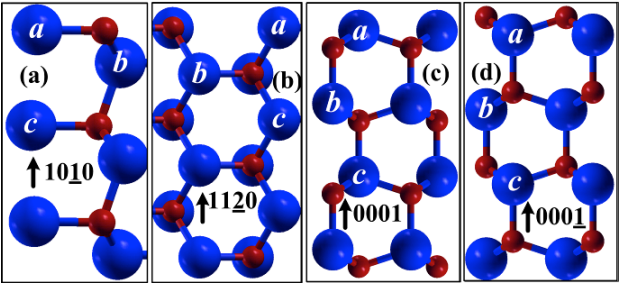

FIG. 1. (Color online) Schematic representation of the side view of perfect surface models: (a) (10$\underline{1}$0), (b) (11$\underline{2}$0), (c) (0001)-Zn, and (d) (000$\underline{1}$)-O surfaces. Zn: blue, O: red. $a$, $b$, and $c$ denote atomic layers of three possible substitutional sites.

Recent studies have shown that the (0001)-Zn polar surface can be reconstructed with about a quarter of the Zn ions missing from the topmost layer [22,23]. Therefore, we have employed corresponding models to investigate vacancies on polar surfaces. The defective (2×2) polar surface models involve ¼ zinc vacancies and ¼ zinc vacancies combined with ¼ oxygen vacancies, hereafter, denoted as $Zn_{0.75}O_{1.00}$ and $Zn_{0.75}O_{0.75}$, respectively. The zinc and oxygen vacant sites ($V_{Zn}$ and $V_O$) have been placed in three different atomic layers (cf. Fig. 2).

It is difficult to define a surface energy of the ZnO polar surfaces due to their nonequivalent terminations. Instead, the cleavage energy, $\gamma_c$, the energetic difference of ZnO slab models ($E_{slab}$) with respect to ZnO bulk ($E_{bulk}$) with identical stoichiometry, is well defined, and employed here to directly compare the stabilities of both polar and non-polar surfaces (see Eq. 1). The doping energy, $\gamma_d$, is defined as difference of the cleavage energy (Eq. 2), and compares the stability of Fe-ZnO with respect to pure ZnO [24]. A negative value of $\gamma_d$ accounts for surface stabilization.

$$\gamma_c = \frac{1}{A}(E_{slab} - E_{bulk}) \quad (1)$$

$$\gamma_d = \gamma_{c(Fe-ZnO)} - \gamma_{c(ZnO)} \quad (2)$$

where $\gamma_{c(Fe-ZnO)}$ and $\gamma_{c(ZnO)}$ denote the cleavage energy of the Fe-ZnO and pure ZnO surface models with respect to surface area (A), respectively. We have converged the thickness of slab models with respect to the cleavage energies. The optimized slabs consisted of 16 layers for (10$\underline{1}$0), 12 layers for (11$\underline{2}$0), and 18 layers for (0001)-Zn and (000$\underline{1}$)-O surfaces.

Following the well-tested protocol of our earlier work [8,9], we have performed density-functional theory (DFT) calculations with the PBE0 hybrid functional [25] as implemented in the CRYSTAL09 code [26]. Full geometry optimization of atomic positions and cell parameters was performed for all studied models. All-electron basis sets were employed (Zn: 86-411(41d)G [27], O: 6-31(1d) [28], Fe: 86-411(41d)G [29]) in order to be independent of parameterization of pseudopotentials. The XANES calculations were carried out using the FEFF9.0 code [30] on the basis of multiple scattering schemes and the core-hole interactions are considered.

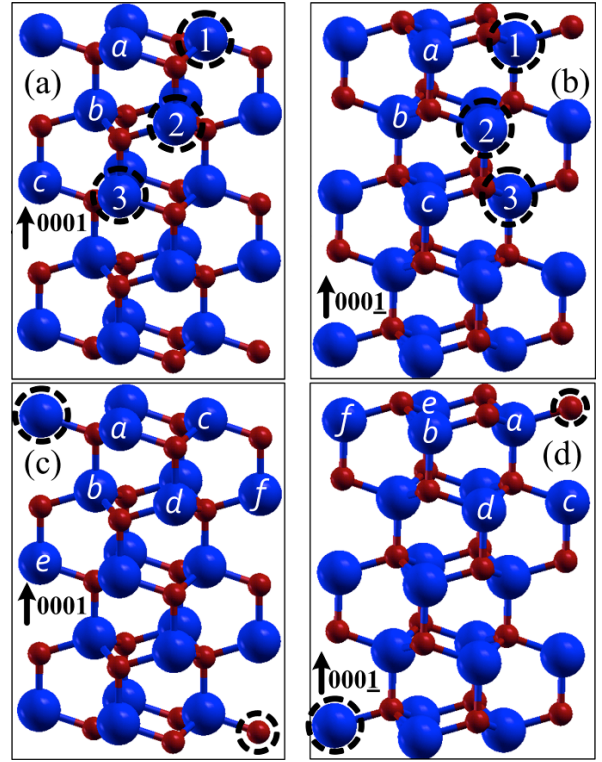

FIG. 2. (Color online) Side views of defective polar surface models are schematically shown in a (2×2) unit cell. (a) $Zn_{0.75}O_{1.00}$-(0001)-Zn, (b) $Zn_{0.75}O_{1.00}$-(000$\underline{1}$)-O, the dashed circles represent three possible vacant sites; (c) $Zn_{0.75}O_{0.75}$-(0001)-Zn, (d) $Zn_{0.75}O_{0.75}$-(000$\underline{1}$)-O surfaces, the dashed circles account for vacant sites. Zn: blue, O: red. The letters denote possible substitutional sites.

The calculated cleavage energies of pure ZnO polar and non-polar surfaces are in good agreement with previous reports [24]. For the non-polar (10$\underline{1}$0) and (11$\underline{2}$0) surfaces we obtained $\gamma_c$ of +0.17 and +0.18 eV Å$^{-2}$, respectively. In contrast, the $\gamma_c$ of the two polar surfaces are much higher (+0.34 eV Å$^{-2}$), confirming that the ZnO polar surfaces are much less stable.

The calculated doping energies of polar and non-polar surfaces are shown in Table I. Positive values of $\gamma_d$ in the case of Fe@a on the (10$\underline{1}$0) and (11$\underline{2}$0) surfaces indicate that Fe should not distribute on non-polar surfaces. This is in agreement with previous experiments, i.e. Fe dopants distribute homogenously in the ZnO NPs without Fe atoms aggregated on the NP surfaces [7,8]. On the other hand, the $\gamma_d$ values of Fe@a on the (000$\underline{1}$)-O and (0001)-Zn surfaces are negative, what demonstrates that the Fe dopants stabilize perfect polar surfaces.

**TABLE I.** Calculated doping energies (in eV Å$^{-2}$) of perfect and defective surfaces following Eqs. 1 and 2.

| Surface | Fe@$a$ | Fe@$b$ | Fe@$c$ |
|---|---|---|---|
| (11$\underline{2}$0) | +0.05 | +0.06 | +0.07 |
| (10$\underline{1}$0) | +0.08 | +0.08 | +0.09 |
| (0001)-Zn | −0.04 | −0.04 | −0.05 |
| (000$\underline{1}$)-O | −0.06 | −0.08 | −0.12 |
| $Zn_{0.75}O_{1.00}$ | Fe@$a$ | Fe@$b$ | Fe@$c$ |
| (0001)-Zn | +0.23 | +0.23 | +0.25 |
| (000$\underline{1}$)-O | −0.05 | +0.26 | +0.23 |
| $Zn_{0.75}O_{0.75}$ | Fe@$a$ | Fe@$b$ | Fe@$c$ |
| (0001)-Zn | −0.27 | +0.65 | +0.02 |
| (000$\underline{1}$)-O | +0.01 | +0.01 | +0.01 |
| $Zn_{0.75}O_{0.75}$ | Fe@$d$ | Fe@$e$ | Fe@$f$ |
| (0001)-Zn | −0.05 | +0.02 | +0.02 |
| (000$\underline{1}$)-O | +0.01 | +0.23 | −0.07 |

We have observed similar stabilization of polar surfaces with various defects, the most prominent being the stabilizations of Fe@$a$ on the $Zn_{0.75}O_{0.75}$-(0001)-Zn (cf. Table I). These results show that stabilization of perfect or defective polar surfaces plays a major role in the experimentally observed stabilization of Fe-ZnO NPs [7].

The selective stabilization of ZnO surfaces by Fe dopants is a very critical feature. In a nanoparticle, a small portion of instable polar surfaces is always present at the edges and vertices. Stabilization of these surfaces results in a reduced number of sites where solvent molecules can successfully attack the nanoparticle and therefore, in lower solubility.

The stabilization of ZnO polar surfaces could be achieved by saturating surface dangling bonds. In the following, we present evidence that this surface saturation can be achieved rather by changes of Fe oxidation states. The availability of an alternative Fe oxidation state, namely $Fe^{3+}$, allows the internal saturation of dangling bonds on polar surfaces. While Fe dopants are preferable present as $Fe^{2+}$ in the ZnO bulk [8], the calculated XANES of surface models reveal that Fe prefers the $Fe^{3+}$ form at the topmost surfaces (cf. Fig. 3, Fe@$a$). Moreover, $Fe^{2+}$ is always present at the subsurface (Fe@$b$) and close to the bulk (Fe@$c$).

One could imagine a simplified chemical process to understand the presence of $Fe^{3+}$ dopants. In the bulk, Fe-ZnO NPs contain mostly $Fe^{2+}$ cations [9]. Once a NP starts to dissolve, Fe atoms arrive at the topmost surface layer. In the case of the (000$\underline{1}$)-O surface, Fe transfers one $d$-electron to saturate the newly formed dangling surface bond, resulting in the $Fe^{3+}$ cation with more stable half-filled $d$-orbital. In contrast, the dangling bonds on non-polar surfaces can be compensated with formation of Zn and O pairs and additional Fe dopants can destroy such a charge balance, resulting in destabilization of ZnO NPs. Therefore, we attribute the observed reduced dissolution of ZnO NPs to the stabilization of polar surfaces.

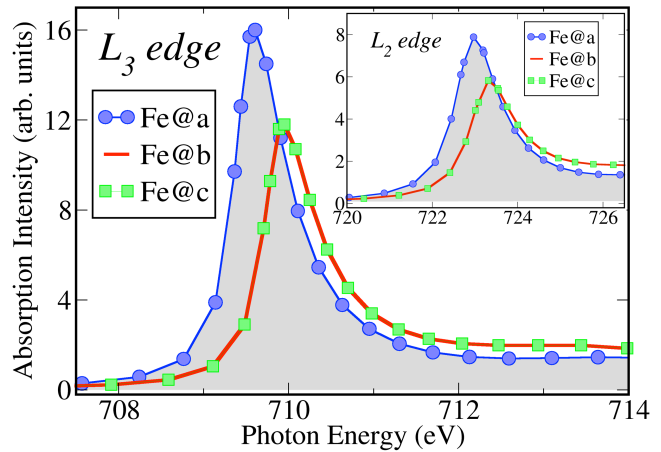

FIG. 3 (Color online) Computed XANES of Fe dopants on the (000$\underline{1}$)-O surface: (a) $L_3$ and (b) $L_2$ edge. For the labeling of models see FIG.1.

Further enhancement of stability is achieved if the macroscopic dipole moment is reduced. To study this effect, we have used combined surface models, where the ZnO bulk has homogeneous distribution of $Fe^{2+}$ atoms [7,8]. Thus, we could calculate the charge transfer in-between two polar surfaces. The total dipole moment of ZnO NPs can be described in terms of charge density of all the layers in the bulk, $\sigma$, and the charge density of the surfaces, $\sigma$'. The dipole moment could be completely cancelled out by a charge transfer in-between two polar surfaces in case when charge densities hold the equation $\sigma$'=76.5%×$\sigma$ [31]. For pure ZnO, the computed charge density of polar surface models is 83.3% $\sigma$, while it is reduced to 76.1% $\sigma$ in the case of Fe-ZnO. Apparently, charge transfer assisted by the Fe dopants is much closer to the required one as compared with pure ZnO polar surfaces. Therefore, the stability of ZnO NPs can be enhanced by suppression of the total dipole moment with Fe doping. The calculated electronic structure serves to interpret the mechanism of the charge transfer assisted by Fe dopants.

The computed O $K$-edge XANES of different surface models in comparison with ZnO bulk are shown in Fig. 4. In the Fe-ZnO NPs with Fe mixed valence states ($Fe^{3+}$ and $Fe^{2+}$), the 3$d$ electrons are directly interacting with O-2$p$ electrons, what activates the electron hopping from 3$d$ orbitals of one Fe to the neighboring one. This is, indeed, in good agreement with the double exchange mechanism proposed by Zener [32]. An $Fe^{2+}$-$Fe^{3+}$ pair can facilitate electron hopping in magnetic systems. The O anion, as a connecter in-between the $Fe^{2+}$ and $Fe^{3+}$ cations, must contribute some unoccupied 2$p$ states. The pre-edge peak of the O $K$-edge in the Fe-ZnO systems indicates the presence of additional electronic states compared with pure ZnO (cf. Fig.4). Comparison between Fe-ZnO and pure ZnO indicates that the peak can be attributed to the partially vacant *d*-orbitals, which are formed due to the hybridization between O-2*p* and Fe-3*d* states [33,34,35]. In other words, the O-2*p* orbitals show a spin polarization that is induced by Fe dopants (cf. Fig.4).

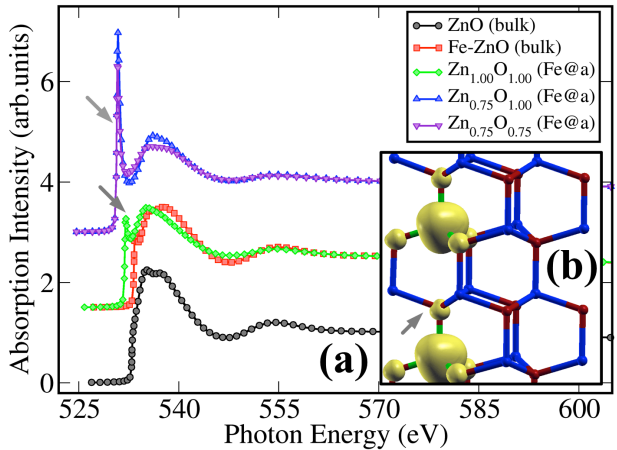

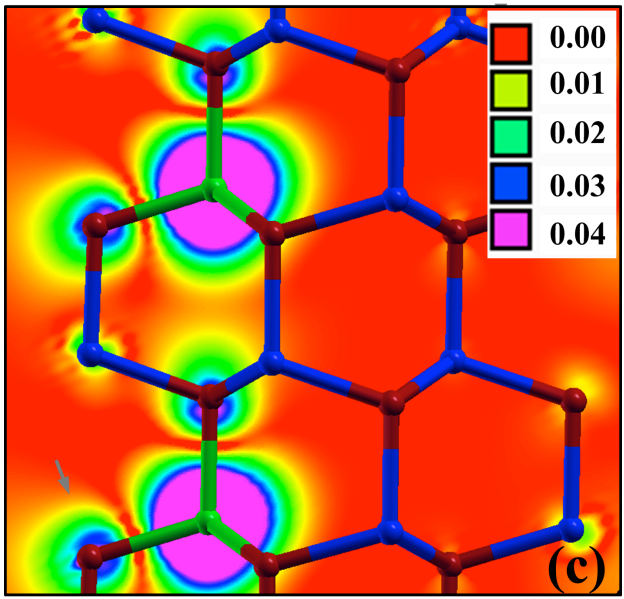

FIG. 4 (Color online) (a) Computed XANES for *K* edge of O atoms in the case of pure ZnO bulk, Fe-ZnO bulk, perfect $Zn_{1.00}O_{1.00}$, as well as defective $Zn_{0.75}O_{1.00}$ and $Zn_{0.75}O_{0.75}$ surfaces. Total spin charge densities ($\rho_{up}$-$\rho_{down}$): (b) isosurfaces at the value of 0.01 $e$ Å$^{-3}$ and (c) isolines at the range from 0 to 0.04 $e$ Å$^{-2}$ and Fe: green, Zn: blue, O: red.

The partial charges of oxygen on the topmost layer of the (0001)-O surface can be transferred effectively to the (000$\bar{1}$)-Zn surface on the opposite side via the hybridized 2*p*-3*d* states. As a result, this process can suppress effectively the total dipole moment and then stabilize polar surfaces indirectly by charge transfer.

In summary, we have investigated the stability of perfect polar and non-polar surfaces in ZnO NPs doped with Fe. Defective polar surfaces with zinc and oxygen vacancies were considered as well. The stabilization by Fe dopants occurs uniquely on polar surfaces, where the dipole moments are strongly suppressed. At the same time, the non-polar surfaces are kept intact. This indicates that the experimentally observed solubility reduction in ZnO NPs is solely attributed to the stabilization of high-energy polar surfaces. Our results show that two stabilization mechanisms are possible: (i) Fe dopants assisted charge transfer suppresses the total dipole moment of ZnO NPs and (ii) the $Fe^{3+}$ cations saturate dangling bonds and thus stabilize the ZnO NPs.

J. Xiao would like to acknowledge financial support from the China Scholarship Council (CSC). The authors thank the Computational Laboratory for Analysis, Modeling, and Visualization (CLAMV) for computational support.